\documentclass[
    ,final            
  ]
  {aipproc}

\usepackage{amsmath}
\layoutstyle{6x9}

\def\lsim{\stackrel{\scriptstyle <}{\phantom{}_{\sim}}}

\begin{document}

\title{Stopping and the $\langle K\rangle/\langle \pi \rangle$ horn}


\author{Boris Tom\'a\v sik}{
  address={The Niels Bohr Institute, Blegdamsvej 17, 2100 Copenhagen \O, Denmark},
  ,altaddress={\'Ustav jadern\'e fyziky AV\v CR, 25068 \v Re\v z, Czech Republic},
  ,email={boris.tomasik@cern.ch}
}

\author{Evgeni E.\ Kolomeitsev}{
  address={School of Physics and Astronomy, University of Minnesota,
           116 Church Street S.E., Minneapolis, Minnesota 55455, USA}
}

\begin{abstract}
We propose a non-equilibrium hadronic model to interpret the
observed excitation function of kaon-to-pion ratios in nuclear
collisions at AGS and SPS energies. The crucial assumption of our
model is that due to stronger stopping at lower energies the
lifetime of the fireball is prolonged because the system has to
build up the longitudinal expansion from internal pressure.
\end{abstract}

\maketitle


\section{The horn}

The excitation function of the ratio
$\langle K^+\rangle / \langle \pi^+ \rangle$ which exhibits 
a sharp peak (a ``horn'')
at projectile energies of 30~$A$GeV is certainly one
of the most intriguing experimental results from the energy
scan performed at the CERN SPS \cite{lung}. Statistical hadronisation
model expects a maximum connected with the transition from baryon
to  meson dominated energy regime but cannot reproduce the observed
sharpness of the peak \cite{Cleymans:2004hj}.
The sharp maximum together with the excitation functions of
$\langle K^- \rangle/\langle \pi^-\rangle$ and
$\langle \Lambda \rangle / \langle \pi \rangle$ is not reproduced
in hadronic transport codes either \cite{bleicher,cassing}.
The only successful, though schematic, interpretation of the data uses
the framework of the so-called 
{\em Statistical model of the early stage (SMES)}
\cite{smes}. The model is based on the assumption that the primordial
particle production is realized according to grand-canonical equilibrium
distribution and the steep descent of the peak corresponds to
transition to deconfined phase via a mixed phase. A kinetic description
of the excitation function, also including phase transition in the
region of the peak, has been proposed recently \cite{jane}.

Concerning the SMES, a question appears whether it is realistic to
assume a chemical equilibrium thus early in the collision? Moreover, if the
peak is to be regarded as a signature for deconfinement, are all
hadronic scenarios safely excluded? We shall explore the
possibility of reproducing the ``horn'' in a hadronic non-equilibrium model.

Let us recall the two important quantities which regulate the final
amount of produced strangeness. Firstly, it is the energy density, because
the rates of reactions producing strangeness
depend on the energy of particles in the incoming channel.
Secondly, it is the total lifespan of the fireball since strange
particles are out of chemical equilibrium and the relative strangeness
content, most probably, grows with time. Thus we propose the following
scenario potentially leading
to the $\langle K^+\rangle / \langle \pi^+ \rangle$ ``horn'':
at lower energies ($\lsim 30A{\rm GeV}$) the increase of the excitation 
function
is due to an increase in the energy available for the strangeness production.
With  the further increase of the collision energy
the nuclear stopping power decreases. Since we know from
$M_T$-dependence of HBT radii that the longitudinal expansion pattern
at freeze-out
looks roughly the same at all energies, there must be accelerated
longitudinal expansion and it must last longer at lower energies with stronger
stopping. We put the sharp decrease of the ratio (on the right-hand side of the
``horn'') into connection with shorter lifetime of the fireball
in more energetic collisions. The fireball just has less time to ``cook up''
strangeness and therefore the relative strangeness yield is lower.
This hypothesis will be tested with a kinetic model.


\section{The model}

As we only want to calculate the ratios of total yields, and not the yields
themselves, it is enough to study the evolution of the 
{\em spatially averaged densities} of individual species only.
The spatial distribution is inessential here. We will in particular
investigate the evolution of {\em kaon} density.
Assuming that kaons are kept in thermal equilibrium with a
fireball medium, the variation of the kaon density follows from the simple
relation
\begin{equation}
\label{dder}
\frac{dn_K}{d\tau} = \frac{d}{d\tau}\frac{N_K}{V} =
- \frac{N_K}{V}\, \frac{1}{V}\, \frac{dV}{d\tau}
+ \frac{1}{V}\, \frac{dN_K}{d\tau}\, ,
\end{equation}
where $N_K$ and $V$ are the total number of kaons and the volume of the
system, respectively, and $\tau$ is the time in the co-moving frame.
The second term on the right-hand side of this equation expresses the
change of density due to chemical reactions. It can be split into
gain term and loss term
\begin{equation}
\label{me}
\frac{dn_K}{d\tau} = n_K \left ( -\frac{1}{V}\, \frac{dV}{d\tau}\right ) +
\sum_{ij} \langle v_{ij}\sigma^+_{ij} \rangle  \frac{1}{1+\delta_{ij}}
n_i\, n_j -
\sum_{j} \langle v_{Kj}\sigma^-_{Kj}  \rangle \frac{1}{1+\delta_{Kj}} n_K\, n_j
\, .
\end{equation}
In the gain term we sum over all processes producing a kaon, in the
loss term these are the processes destroying a kaon. The
relative-velocity-averaged cross-sections are multiplied with
densities of the incoming species.

The first term on the right-hand-sides of eqs.~\eqref{dder} and \eqref{me}
includes the expansion rate and
stands for the density change due to expansion. In a simulation of the
collision (hydrodynamic or transport), the expansion rate is obtained
naturally. Here, we shall adopt a {\em parametrisation}
of the expansion. Though in this way we do not make
a direct contact to the underlying microscopic
equation of state, it gives us a possibility to explore many different
evolution scenarios and their impact on data.

The ansatz for the energy density and the baryon density which will
be used reads
\begin{subequations}
\label{pars}
\begin{eqnarray}
\label{tpar}
\varepsilon(\tau) & = &
\begin{cases}
\varepsilon_0 (1 - a\tau - b\tau^2)
& \mbox{for} \quad \tau < \tau_s \\
\frac{\varepsilon^\prime_0}{(\tau - \tau_0)^{\alpha/\delta}} & \mbox{for} \quad
\tau > \tau_s
\end{cases}\\
\rho_B(\tau) & = &
\begin{cases}
\rho_{B,0} (1 - a\tau - b\tau^2)^\delta
& \mbox{for} \quad \tau < \tau_s \\
\frac{\rho_{B,0}^\prime}{(\tau - \tau_0)^{\alpha}} & \mbox{for} \quad
\tau > \tau_s
\end{cases}
\end{eqnarray}
\end{subequations}
In the above expressions the parameters can be tuned. Two of them
are constrained by the requirements that the functions
are continuous together with their first time derivatives.
Initially ($\tau<\tau_s$) the fireball expands acceleratedly
with a quadratic dependence on time. Then, at $\tau>\tau_s$,
the power-law expansion is dictated by the $M_\perp$-dependence of HBT
radii. Note the shift by $\tau_0$: it allows to delay this type of
expansion. Such a delay would be unobservable
in the $M_\perp$-dependence of $R_{\rm long}$ which is standardly used
to argue for a short fireball lifetime~\cite{cereshbt}. 
The ``lifetime'' measured by
$R_{\rm long}$ would correspond in our case to the difference
$(\tau_{\rm total}-\tau_0)$. Note also that the two parameterisations
for energy and baryon density have similar forms and differ essentially
by the exponent $\delta$ which is given by the assumed equation
of state.

By tuning the parameters $a$ and $b$ we specify the initial rate of density
decrease and the pace on which the decrease accelerates. This way we
quantify the stopping and re-acceleration. We can construct a large class
of models ``between Landau and Bjorken scenarios''.

Densities of kaons (i.e.\ $K^+,\, K^0,\, K^{*+},\, K^{*0}$) are calculated
according to eq.~\eqref{me}. We include the following processes:
\begin{align*}
& {\pi N \leftrightarrow K Y}, && \pi N \to NK\bar K, &&
{\pi \Delta \leftrightarrow KY},&& \pi\Delta \to NK\bar K\\
& NN \to KNY,&& NN \to NNK\bar K,&& NN \to \Delta KY && \\
&N \Delta \to NYK,&& N\Delta \to NNK\bar K,&& N \Delta \to \Delta KY\\
&\Delta \Delta \to \Delta YK,&& \Delta\Delta \to NNK\bar K\\
&{\pi\pi \leftrightarrow K\bar K},&&
{\pi \rho \leftrightarrow K\bar K}, &&
{\rho\rho \leftrightarrow K\bar K}, && \\
&{K^* \leftrightarrow K \pi}, &&
{\pi Y \leftrightarrow K \Xi}. && &&
\end{align*}
For those processes with two particles in final state also
inverse reactions have been included which annihilate kaons.

Species with negative strangeness must balance the total strangeness
of the system to zero. Reactions which just swap the strange quark
between them are quick. Therefore, we assume that these species are
in chemical equilibrium with respect to each other, while the
strangeness-weighted sum of their densities is given by the requirement
of strangeness neutrality.

All non-strange species are assumed to be chemically equilibrated. We assume
no antibaryons at these energies. Their influence would be highest
at the top SPS energy $\sqrt{s_{NN}} = 17\, A\mbox{GeV}$ where we
expect 10\% error due to their neglect. In a schematic model like ours
this is acceptable.

In a model set up in this way, the total produced amount of strangeness
is mainly controlled by the lifetime (as we shall show) and slightly
less by the temperature. The rate at which strange quarks are distributed
among $K^-$ and $\Lambda$ is fixed by the final temperature.

We try to choose the parameterisations \eqref{pars} in such a way that
the final state resulting from fits to chemical composition at different
energies \cite{beca}
is reached. The initial amount of strangeness is estimated from
pp, pn, and nn collisions \cite{inprep}.


\section{Results}

\begin{figure}
  \includegraphics[height=.4\textheight]{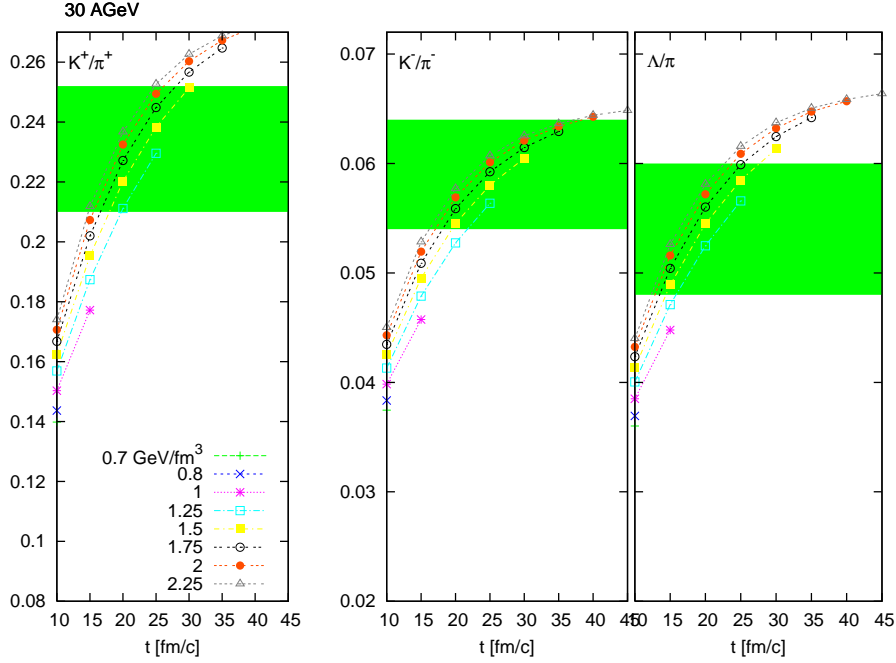}
  \caption{Calculated ratios $\langle K^+\rangle /\langle \pi^+\rangle$,
    $\langle K^-\rangle /\langle \pi^-\rangle$,
    $\langle \Lambda\rangle /\langle \pi\rangle$ as functions of the
    total lifetime of the fireball. Different curves correspond to different
    initial energy densities. Calculation for Pb+Pb collisions at projectile
    energy of 30~$A$GeV. The bands show values accepted by data \cite{lung}.
  \label{f:res}}
\end{figure}

We show an example of our results in Figure~\ref{f:res}.
Each point in
that figure corresponds to a different evolution scenario; they differ
by total lifetime and initial energy density. Though the presented
results are obtained for Pb+Pb collisions at projectile energy of 30~$A$GeV,
it applies generally that dependence on lifetime is crucial, while
the dependence on initial energy density is less important. The latter
gains more weight at lower energies and is completely irrelevant at the
highest SPS energy. It mainly stems from the temperature dependence of
reactions including nucleons like $\pi+N\to \Lambda + K$ which make up a
bigger share of the total production rate at lower energies where the baryon
density is higher.

\begin{figure}
  \includegraphics[height=.4\textheight]{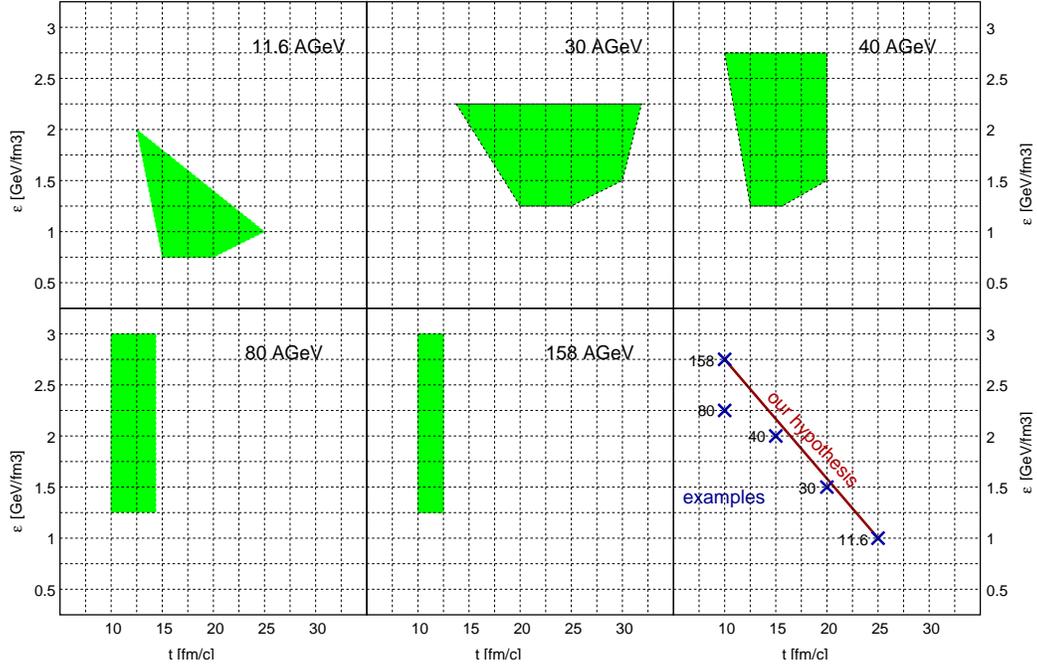}
  \caption{Values of initial energy density and total lifetime accepted
    by data \cite{lung}.
  \label{f:accept}}
\end{figure}
\begin{figure}
  \includegraphics[height=.4\textheight]{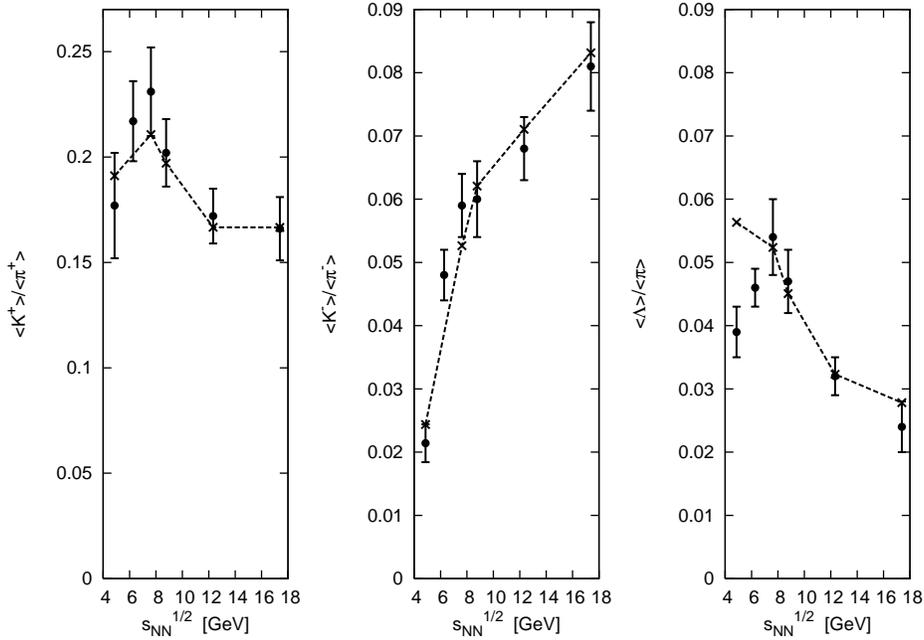}
  \caption{Results obtained with parameter sets indicated
    in lower right panel of Figure~\ref{f:accept} compared to
    data.
  \label{f:comp}}
\end{figure}

%
Summary of the values of total lifetime and initial energy density allowed
by comparison to data is presented in Figure~\ref{f:accept}. The more important
limitations are put on the lifetime. In general, for SPS energies
our hypothesis is confirmed that with increasing energy of the collision
lifetime of the fireball decreases. Only at the lowest studied energy,
11.6~$A$GeV (AGS), we miss the $\langle \Lambda\rangle / \langle \pi \rangle$
ratio by about 3 standard deviations, see Figure~\ref{f:comp}.
%
This is caused by much too high temperature in the final state, which
can be fixed by re-parameterising the time dependence of the energy density.
Otherwise, we reproduce the data well.


\section{Conclusions}

The excitation function of the  ratios
$\langle K^+ \rangle/\langle \pi^+\rangle$,
$\langle K^- \rangle/\langle \pi^-\rangle$ and
$\langle \Lambda \rangle / \langle \pi \rangle$ can be reproduced
in a non-equilibrium hadronic scenario. The crucial assumption
is that the total lifetime of the fireball is a decreasing
function of the collision energy.

Thus we have made a specific prediction for the evolution dynamics
of the fireball. Such a prediction must now be checked against other
observables. Although we were lead by the knowledge of all particle
abundances, transverse momentum spectra and HBT radii, a careful
analysis of these data in the framework of the proposed
model must be performed. Furthermore, dilepton spectra
gained an unprecedented accuracy recently \cite{ceres,ceres40,na60}. They are
sensitive to whole fireball history \cite{renk} and could provide
the crucial test of our hypothesis.


\section{Acknowledgements}

The work of BT was supported by a Marie Curie Intra-European
Fellowship within the 6th European Community Framework Programme.
The work of EEK was supported by the US Department of Energy
under contract No. DE-FG02-87ER40328.

\end{document}